\documentclass[11pt]{article}

\global\arraycolsep=1pt

\usepackage{amsmath}
\usepackage{amsfonts}
\usepackage{bbm}
\usepackage{graphicx}
\usepackage{epsf}
\newcommand{\ben}{\begin{eqnarray}}
\newcommand{\een}{\end{eqnarray}}
\newcommand{\bee}{\begin{equation}}
\newcommand{\eee}{\end{equation}}
\newcommand{\beo}{\begin{equation*}}
\newcommand{\eeo}{\end{equation*}}

\newcommand{\bb}[1][\overline]{1}

\newcommand{\IZ}{\mathbb Z}
\newcommand{\IC}{\mathbb C}
\newcommand{\IR}{\mathbb R}

\def\x{a}

\DeclareMathOperator{\CP}{\mathbb{P}}

\def \x{\tilde{x}}
\def \at {\hat{t}}
\def \tiP{\tilde{P}}
\def \tiQ {\tilde{Q}}
\def\hl{\hat{l}}

\def\braket#1#2{{\langle#1,#2\rangle}}
\DeclareMathOperator{\gcf}{g\!\;\!c\!\;\!f}

\long\def\symbolfootnote[#1]#2{\begingroup%
\def\thefootnote{\fnsymbol{footnote}}\footnote[#1]{#2}\endgroup}

\begin{document}

\begin{titlepage}

\begin{flushright}
\normalsize
~~~~
LMU-ASC 28/11\\ CERN-PH-TH/2011-166

\end{flushright}

\vspace{15pt}

\begin{center}
{\LARGE  Reflections of NS5 branes} \\

\end{center}

\vspace{23pt}

\begin{center}
{ Adrian Mertens\symbolfootnote[1]{adrian.mertens@physik.uni-muenchen.de}
}\\
\vspace{20pt}

\it Arnold Sommerfeld Center for Theoretical Physics\\
LMU, Theresienstr. 37, D-80333 Munich, Germany
\vspace{5pt}

\it Department of Physics, CERN Theory Divison\\
CH-1211 Geneva 23 Switzerland

\vspace{5pt}
\end{center}
\vspace{20pt}
\begin{center}
Abstract\\
\end{center}

We study complex structure monodromies of certain Calabi-Yau fibrations and find evidence that they are mirror to Calabi-Yau manifolds with NS5 brane on a divisor. This gives a simple way to construct mirrors to any Calabi-Yau hypersurface with NS5 branes wrapped on divisors and a complementary interpretation of some recent calculations in open string mirror symmetry.


\vfill

\end{titlepage}

\section{Introduction}

The equivalence of NS5 branes and certain Ricci flat geometries under T-duality was first shown in \cite{Ooguri:1995wj} by a study of the conformal field theory for ALE spaces with $A_{N-1}$ singularities. These geometries are $S^1$ fibrations with $N$ vanishing fibers. A T-duality along the fiber turns it into $N$ parallel NS5 branes in flat space. The T-duality acts in a normal direction to the resulting NS5 brane, the localization of the brane in this direction is due to instanton effects \cite{Gregory:1997te}. The following geometric explanation based on \cite{Greene:1989ya} was already given in \cite{Ooguri:1995wj}: The effect of an NS5 brane localized on a point in a Torus $Z^*$ and a point $\IC$ is a monodromy of the $B$-field, $B\to B+2\pi$ around the brane in $\IC$. This gives a monodromy $\rho\to\rho+1$ of the complexified K\"ahler class $\rho=\frac{B}{2\pi}+i\sqrt{G}$. Mirror symmetry, or T-duality in one $S^1$ of the torus, exchanges the K\"ahler class with the complex structure. After T-duality one thus expects a monodromy $\tau\to\tau+1$ for the dual torus $Z$. To get such a monodromy the dual geometry has to be a fibration of $Z$ over $\IC$. Instead of an NS5 brane there is a singular fiber with a shrinking $S^1$. 

Mirror symmetry should also geometrize NS5 branes on divisors in higher dimensional Calabi-Yau (CY) spaces \cite{Aganagic:2009jq}. In the Strominger Yau Zaslow picture \cite{Strominger:1996it} one of the T-dualities in the Lagrangian torus fiber acts in a normal direction of a generic divisor. While T-dualities in an internal direction map an NS5 brane to another NS5 brane, the single T-duality in a normal direction should turn it into a locus of a vanishing $S^1$. The resulting fibration has to be a consistent background preserving the same amount of supersymmetry, so it should be a non compact CY space.
Based on this idea we propose a global description of a dual geometry for NS5 branes localized on a point in $\IC$ and a divisor in a n-dimensional CY hypersurface $Z^*$ in a toric ambient space. The dual geometry $X$ is a fibration of the mirror CY $Z$ over $\IC$ and is itself a non-compact n+1 dimensional CY hypersurface. It can be constructed by toric methods. We study the complex structure monodromy of the fibers and find perfect agreement with the dual K\"ahler monodromies created by NS5 branes.

More concretely we propose that specific non-compact CY manifolds that already appeared in \cite{Alim:2009rf,Alim:2009bx,Alim:2010za} can be interpreted in this way. These papers discuss superpotentials for branes wrapped on curves in CY 3-folds $Z^*$. The curves are first immersed into a divisor. Then the unobstructed deformation space of the divisor inside the CY is used to calculate volumes of chains ending on curves within the divisor. These relative period integrals were seen to be equivalent to period integrals of a non compact CY 4-fold $X^*$. The mirror $X$ to this non compact 4-fold is the geometry we will mainly study in this note. The matching of relative period integrals for divisors in $Z^*$ with quantum corrected volumes of cycles in $X$ can be interpreted as first evidence for the present proposal. 
In \cite{Jockers:2009ti} this match was explained by a different chain of dualities starting from 7-branes on the divisor and involving heterotic/F-theory duality. It would be interesting to close both proposals to a cycle of dualities. The present proposal appeared implicitly already in \cite{Aganagic:2009jq}, where a similar construction involving NS5 branes is used to calculate superpotentials. 

The paper is organized as follows: In chapter 2 we explain the construction of the non compact manifold $X$ for a given divisor on a CY hypersurface $Z^*$. In chapter 3 we consider the example of an NS5 brane on a torus. We calculate the monodromy of the complex structure of the fiber in the proposed dual geometry and show that it matches with the shift of the $B$-field. In chapter 4 we take $X$ to be a K3 fibration and study the central fiber in detail. Chapter 5 contains the main observation. The complex structure monodromies for the fibers of $X$ always map to the expected shift of a $B$-field. We use toric methods and the relation between monomials and divisors of mirror manifolds. Chapter 6 contains further examples and chapter 7 some conclusions. We also comment on a "mirror" mapping between the divisor and the degeneration locus. This could be interesting as generically the degeneration locus is not CY.

\section{The dual geometry}
\label{construction}

We start by repeating the construction of the non compact CY fibration $X$, \cite{Mayr:2001xk,Alim:2009rf,Alim:2009bx}. We use standard notation for polytopes in mirror symmetry, see e.g. \cite{mibo}. Some details are summarized in \ref{polytopes}.\newline
A CY hypersurface $Z^*$ in a toric variety is given as the vanishing locus of an equation
\begin{equation*}
\tiP(Z^*)=\sum_i a_i \x^{\nu_i}\,.
\end{equation*}
The monomials $\x^{\nu_i}=\prod_j \x_j^{\nu_{i,j}}$ appearing in this equation are labeled by integral points $\nu_i$ of some reflexive lattice polytope $\Delta^*$. There are relations $\sum_i l_i^a \nu_i=0$ between these points and therefore between the monomials, $\prod_i (\x^{\nu_i})^{l_i^a}=1$. These relations can be used to derive a Picard-Fuchs system for the periods of $Z^*$ and to define the gauged linear sigma model (GLSM) \cite{Witten:1993yc} of the mirror CY $Z$.

In $Z^*$ we study the most general divisor $\mathcal{D}$ of a given divisor class without any rigid component. If the degree of the defining equation $\tiQ=0$ for this divisor is not higher than the degree of $\tiP$, it can be expressed as\footnote{If the degree is higher a straightforward analogous construction is still possible. In this case further new points can be added to the extended polytope. We will not consider this case to avoid cluttering the notation.} 
\begin{equation}
\label{divisor}
\tiQ=(b_1 \x^{\nu_a}+b_2 \x^{\nu_b}+ ...+b_n \x^{\nu_*})/\gcf\,,
\end{equation}
where $\gcf$ is the greatest common factor of the monomials in $\tiQ$. There are new relations between the monomials of $\tiQ$ and the monomials of $\tiP$. They lead to a Picard-Fuchs system governing the volume of chains ending on the divisor $\mathcal{D}$, \cite{Li:2009dz,Alim:2009rf,Alim:2009bx}. For bookkeeping we express them as relations $\sum_i\hat{l}_i^a\hat{\nu}_i=0$ between points $\hat{\nu}_i$ of an enlarged lattice polytope $\hat{\Delta}^*$. To construct it we embed $\Delta^*$ in a lattice with one additional dimension and add one point for every monomial in $\tiQ$, $\hat{\Delta}^*=\{(\Delta^*,0),(\nu_a,1),(\nu_b,1), ...,(\nu_*,1)\}$. In the following we use the notation $\hat{l}$ only for the new relations that involve some of the additional points $(*,1)$. Relations involving only the points $(\Delta^*,0)$ are called $l$.

The GLSM defined by a basis for these relations gives a non compact CY $X$. This is the geometry we will mainly study in the following. It is always an $Z$ fibration over $\IC$ with a single singular fiber. In the singular fiber an $S^1$ shrinks over a codimension two locus. As we will show in the following, the complex structure monodromy of $Z$ around this singular fiber matches the $B$-field monodromy of $Z^*$ for an NS5 brane wrapped on $\mathcal{D}$. Moreover the relative periods of the pair $(Z^*,\mathcal{D})$ are mirror to cycles of $X$, including quantum corrections. In particular the moduli of the divisor $\mathcal{D}$ are mapped to K\"ahler moduli controlling the location of the shrinking $S^1$ in the singular fiber.

We thus conjecture that the geometry $X$ is dual to $Z^*\times \IC$ with an NS5 brane wrapped on the divisor $\mathcal{D}$ and localized at a point in $\IC$. 
This statement appeared implicitly already in \cite{Aganagic:2009jq}.\footnote{There an NS5 brane on a divisor is geometrized by a T-duality and the resulting geometry is used to calculate superpotentials. Instead of $Z^*\times \IC$ \cite{Aganagic:2009jq} starts with a CY 3-fold $Z^*$ times $S^1\times \IR$ and performs a T-duality on the $S^1$ to get a 4-fold $Y$ without branes. It was noted that 3 dimensional mirror symmetry of $Z^*$ should also geometrize the NS5 brane and that the resulting geometry could be the (4 dimensional) mirror of $Y$. On the level of period integrals the mirror symmetry between $Y$ and $X$ was checked. $Y$ is however not identical with the mirror $X^*$ of the non compact CY $X$ as it appeared in \cite{Alim:2009rf,Alim:2009bx}. The pair $(X,X^*)$ can be compactified to a mirror pair of compact CY hypersurfaces.}
The internal directions of the NS5 brane fill the divisor in $Z^*$ and the remaining unspecified directions, the dimension of $Z^*$ does not matter. 
By mirror symmetry, instanton effects for $X$ are naturally captured by the classical geometry $X^*$ or the pair $(Z^*,\mathcal{D})$. From a supergravity point of view they cause the localization of the NS5 brane in the transverse circle \cite{Gregory:1997te,Tong:2002rq}, for a recent discussion within "doubled geometry" see \cite{Jensen:2011jn}.

\begin{figure}[!htb]
\centering
\includegraphics[width=12cm]{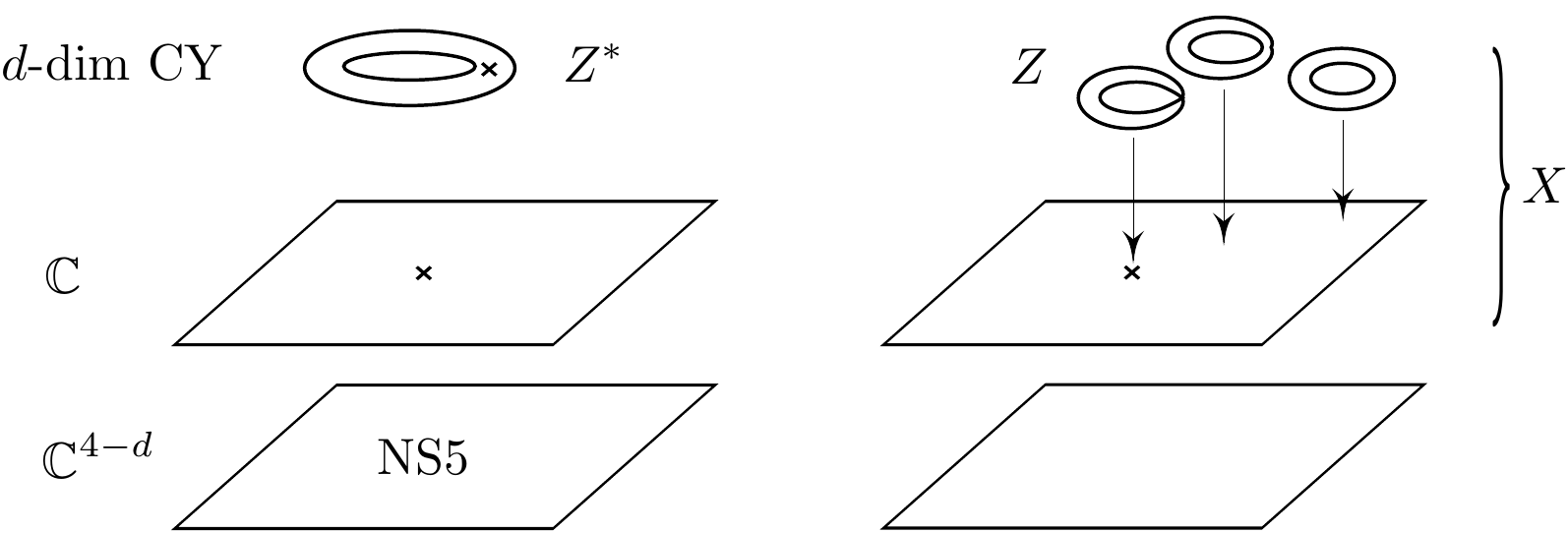}

\vspace{0.5cm}
An NS5 brane localized on a divisor and a point in $\IC$, and the dual geometry $X$.
\end{figure}

\section{NS5 brane on a Torus}

Consider a torus $Z^*$, defined by a hypersurface equation $\tiP=a_1\x_1^3+a_2\x_2^3+a_3\x_3^3+a_0 \x_1\x_2\x_3=0$ in $\mathbb{P}^2/\mathbb{Z}_3$. To this geometry we add an NS5 brane at $\tiQ=b_1 \x_1^2+b_0 \x_2\x_3=0$, localized at the origin of $\mathbb{C}$ and wrapping $\IR^6$. If we forget about the NS5 brane and apply mirror symmetry to the torus we get a dual Torus $Z$. From \cite{Ooguri:1995wj} we learn that we get a fibration of the dual torus $Z$ over $\IC$ if we take the NS5 brane into account.
In the case at hand we wrap an NS5 brane around a divisor of class $2[pt]$.\footnote{This is the simplest example, we will explain later how to construct the dual geometry as well for a Torus with NS5 brane on a minimal divisor.} As the class of a point, $[pt]$, is dual to the K\"ahler class, this gives a monodromy $\rho\to\rho+2$.  We conjecture that the dual non-compact CY 2-fold is given by the GLSM
\begin{center}
\bee\label{Torus1mod}\phantom{arrrgh}\eee\vskip-1.5cm
\begin{tabular}{cc|cccccc}
& $P$ & $x_1$ & $x_2$ & $x_3$ & $y_0$ & $y_1$ \\
\hline
$l$ & -3& 1 & 1 & 1 & 0 & 0  \\
$\hat{l}$ & -1 & 1 & 0 & 0 & 1 & -1 \\
\end{tabular} .
\end{center}
The most general hypersurface equation for these charges is
\begin{equation*}
P=x_1p^2(x_1y_1,\,x_2,\,x_3)+ y_0\, q^3(x_1y_1,\, x_2,\, x_3) + \mathcal{O}(y_0y_1)\,,
\end{equation*}
where $p^2$ and $q^3$ are arbitrary degree two and three polynomials in $x_1y_1$, $x_2$ and $x_3$.
This geometry is a $Z$ torus fibration over $\IC$, the coordinate on $\IC$ is $y_0y_1$. We can see this as follows. $\{(x_1,x_2,x_3)\in \CP^2 | P(x_1,x_2,x_3,y_0,y_1)=0\}$ is a torus whose complex structure depends on $y_0$ and $y_1$. By the D-term constraint for the charge vector $\hat{l}$, $|x_1|^2+|y_0|^2-|y_1|^2=\hat{t}$, the two coordinates $y_0$ and $y_1$ are not independent. We can use this constraint together with the corresponding $U(1)_{\hat{l}}$ action to fix $y_0$ and $y_1$ once the product $y_0y_1\in \IC$ is given. We have thus a torus over each generic point $y_0y_1$ of the base.
The only non generic point is $y_0y_1=0$, where a $S^1$ shrinks in the central fiber. For $|x_1|^2=\hat{t}$ both $y_0$ and $y_1$ vanish and $U(1)_{\hat{l}}$ acts only on the phase of $x_1$. By construction this action is compatible with the hypersurface constraint for the torus at $y_0=y_1=0$, $P=x_1 p^2(x_2,x_3)$. So we can use a cylinder $a<|x_1|<b$ with $a<\sqrt{\hat{t}}<b$ as coordinate patch for the torus and the $U(1)_{\hat{l}}$ action cuts the cylinder into a union of two cones. As there are two solutions to $P=x_1 p^2(x_2,x_3)=0$ with $|x_1|^2=\hat{t}$, this happens twice. The two loci are mirror to the two points $\tiQ=b_1 \x_1^2+b_0 \x_2\x_3=0$. The K\"ahler modulus $\hat{t}$ that determines the position of the degenerating $S^1$ in the torus $T$ is mirror to the modulus $\hat{z}=\frac{a_1 b_0}{a_0 b_1}$ that determines the position of the NS5 branes in $Z^*$. For details on the mirror map in slightly more complicated examples see \cite{Alim:2009rf,Alim:2009bx}.

To calculate the monodromy around the origin we consider $y_0,\,y_1$ as (redundant) parameters that determine the complex structure of the fiber. 
The period integrals can be brought into the standard form by a rescaling $x_1\to x_1 \frac{1}{y_1^{2/3}},\,x_2\to x_2 y_1^{1/3},\,x_3\to x_3 y_1^{1/3}$, 
\begin{equation*}
 \int\frac{\Xi}{x_1p^2(x_1y_1,\,x_2,\,x_3)+y_0\, q^3(x_1y_1,\, x_2,\, x_3)}=\int\frac{\Xi}{x_1p^2(x_1,x_2,x_3)+y_0y_1\, p^3(x_1, x_2, x_3)}\,,
\end{equation*}
where $\Xi$ is the holomorphic 2-form of $\CP^2$. After this rescaling $P$ depends on $y_0$ and $y_1$ only in the combination $y_0y_1$, so we can treat it as hypersurface equation of the fiber depending on the position of the base, $P(x_1,x_2,x_3; y_0y_1)$. The geometry \eqref{Torus1mod} is a blow-up of the fibration $\{(x_1,x_2,x_3)\in\CP^2|P(x_1,x_2,x_3;y_0y_1)=0\}\to\IC$. We discuss this in more detail in the next example.
Close to $y_0y_1=0$ all monomials containing only $x_2$ and $x_3$ are suppressed. After some coordinate redefinitions these are only the two monomials $x_2^3$ and $x_3^3$. Moving in a sufficiently small circle around $y_0y_1=0$ these are the only monomials whose prefactors in the hypersurface equation of the fiber vary and we can use standard methods \cite{Hosono:1993qy} to determine the complex structure. We find $\tau=2 \ln(y_0y_1)+\mathcal{O}(y_0y_1)$ near $y_0y_1=0$. The factor of 2 comes about as both the monomials $x_2^3$ and $x_3^3$ are suppressed by $y_0y_1$. Alternatively, after an additional rescaling, only one of the monomials e.g. $x_2^3$ is suppressed by $(y_0y_1)^2$. This gives the expected monodromy $\tau\to\tau+2$. The logarithmic singularity at $y_0y_1=0$ is in accordance with the expected backreaction of an NS5 brane, for a recent study of this situation in the heterotic string see \cite{Blaszczyk:2011ib, Quigley:2011pv}.

\section{NS5 brane on a K3}
We now want to extend this construction to more complicated geometries. The Strominger Yau Zaslow picture of mirror symmetry \cite{Strominger:1996it} seems to indicate that this is possible. It explains mirror symmetry as simultaneous T-dualities in all directions of a Lagrangian torus fibration. One of this directions is normal, the others transversal to a holomorphic divisor. A T-duality performed in an internal direction maps an NS5 brane to another NS5 brane, a T-duality in a normal direction should turn it into a locus where the T-dual $S^1$ shrinks. Mirror symmetry should thus geometrize the NS5 brane. As it exchanges K\"ahler and complex structure moduli, die shift of the $B$ field that signals the presence of an NS5 brane should be mapped to a monodromy of the complex structure.

We will study generalizations of the geometry \eqref{Torus1mod} and show that the complex structure of the fiber $Z$ has a monodromy around the origin of the base. This monodromy is in agreement with the interpretation as mirrors of a CY $Z^*$ with NS5 brane.
To make contact with the easier case of the torus we consider an elliptically fibered K3 $Z$ that is fibered over $\IC$,
\begin{center}
\bee\label{K3fib}\phantom{arrrgh}\eee\vskip-1.5cm
\begin{tabular}{cc|cccccccc}
& $P$ & $x_1$ & $x_2$ & $x_3$ & $x_4$ & $x_5$ & $y_0$ & $y_1$ \\
\hline
$l^1$ & -3 & 1 & 1 & 1 & 0 & 0 & 0 & 0 \\
$l^2$ & 0 & 0 & 0 & -2 & 1 & 1 & 0 & 0 \\
$\hl$ & -1 & 1 & 0 & 0 & 0 & 0 & 1 & -1 \\
\end{tabular} .
\end{center}
The coordinate on $\IC$ is $y_0y_1$. We call the whole fibration again $X$. It has a singular fiber over the origin of $\IC$, in this singular K3 the elliptic fiber degenerates. Now we concentrate on a neighborhood of the vanishing $S^1$ in the degenerate elliptic fiber. Locally the geometry is a cone ($uv=0$) over $\CP^1 \times \IC$. This should turn into an NS5 brane, if we can consistently implement a duality transformation that involves one T-duality in the elliptic fiber of the K3 $Z$. Mirror symmetry in the $Z$ fiber over each point in $\IC$ is such a duality, its maps the $Z$ fibration to a $Z^*$ fibration over $\IC$. As the K\"ahler structure of $Z$ fiber does not vary in \eqref{K3fib}, the complex structure of $Z^*$ is constant in the dual fibration.

The complex structure of the fiber $Z$ however does vary. The hypersurface equation
\begin{equation}
\label{K3P}
P=x_1 p^2(x_1y_1,x_2,x_3x_4^2,x_3x_5^2,x_3x_4x_5)+y_0 q^3(x_1y_1,x_2,..)+\mathcal{O}(y_0y_1)\,,
\end{equation}
depends on parameters $y_0$ and $y_1$. With $x_i$ we denote coordinates for the smooth blow-up of $\CP_{1122}$. In the following this blow-up is understood whenever we write $\CP_{1122}$ or $\CP_{112}$.
\newline
In the period integrals we can rescale $x_1\to x_1/y_1^{2/3},\, x_2\to x_2 y_1^{1/3},\,x_3\to x_3 y_1^{1/3}$,
\begin{equation*}
 \int\frac{\Xi}{P}=\int\frac{\Xi}{x_1 p^2(x_1,x_2,x_3x_4^2,x_3x_5^2,x_3x_4x_5)+y_0y_1 q^3(x_1,x_2,..)}\,,
\end{equation*}
so the complex structure only depends on the product $y_0y_1$, as it should. Here we claimed that the complex structure of the $Z$ fiber is the same as the complex structure of the hypersurface $P'=x_1 p^2(x_1,x_2,x_3x_4^2,x_3x_5^2,x_3x_4x_5)+z q^3(x_1,x_2,..)+ \mathcal{O}(z^2)$ in $\CP_{1122}$. Let us look at the two geometries more carefully. Both geometries fall apart into two components at $y_0y_1=0$ and $z=0$ respectively. For the fiber $Z|_{y_0y_1=0}$ we have the components
\begin{align*}
|x_1|^2\leq \hat{t}\,,&\;y_1=0 \;\;\text{and}\\
|x_1|^2\geq \hat{t}\,,&\;y_0=0 \,,
\end{align*}
where $\hat{t}$ is the K\"ahler modulus for the charge vector $\hat{l}$, $|x_1|^2+|y_0|^2-|y_1|^2=\hat{t}$.
For the hypersurface $P'=0$ we have
\begin{align*}
 &\{x_1=0\}\in\CP_{1122} \simeq \CP_{112}\;\;\text{and}\\
&\{p^2(x_1,x_2,x_3x_4^2,x_3x_5^2,x_3x_4x_5)=0\}\in \CP_{1122}\,.
\end{align*}
In the first component of the fiber $Z|_{y_0y_1=0}$, $y_1=0$, we have the equation
$x_1 p^2(0,x_2,x_3x_4^2,..)+y_0 q^3(0,x_2,x_3x_4^2,..)=0$. This can unambiguously be solved for $\frac{x_1}{y_0}$ for any ${(x_2:x_3:x_4:x_5 )}\in\CP_{112}$ away from $p^2(x_2,x_3x_4^2,..)=q^3(x_2,x_3x_4^2,..)=0$. Once $\frac{x_1}{y_0}$ is fixed, $x_1$ and $y_0$ are determined by the K\"ahler parameter $\at$. However, at $p^2(x_2,x_3x_4^2,..)=q^3(x_2,x_3x_4^2,..)=0$, the ratio $\frac{x_1}{y_0}$ is free and $(x_1:y_0)$ parameterize a $\CP^1$. So the first component is a $\CP_{112}$, with the locus $p^2(x_2,x_3x_4^2,..)=q^3(x_2,x_3x_4^2,..)=0$ blown up by a $\CP^1$. The size of this $\CP^1$ is the K\"ahler modulus $\at$. In the second component of the fiber, $y_0=0$, we have the equation $x_1p^2(x_1y_1,x_2,x_3x_4^2,x_3x_5^2,x_3x_4x_5)=0$. As $x_1\not=0$ in this component we have $\{p^2(x_1y_1,x_2,x_3x_4^2,x_3x_5^2,x_3x_4x_5)=0\}\in \CP_{1122}$ as for the second component of the hypersurface $P'=0$. The coordinates on $\CP_{1122}$ in this case are $(y_1:x_2:x_3:x_4:x_5)$, so $x_1$ is exchanged for $y_1$. Away from the singular point we have an isomorphism between the K3 fiber and the hypersurface $P'$ by the rescaling given above.\footnote{We see that the singular fiber is a union of two Fano varieties. $Y_1=\{p^2(x_2,x_3x_4^2,x_3x_5^2,x_3x_4x_5)=0\}\in \CP_{1122}$ and $Y_2$ is a blow-up of $\CP_{112}$. They intersect over a Torus $D=\{p^2(x_2,x_3x_4^2,x_3x_5^2,x_3x_4x_5)=0\}\in\CP_{112}$, $K_D=0$ so $D\in|-K_{Y_i}|$. The singular fiber is a normal crossing of the type described in \cite{Lee:2010xw}, while the whole non-compact 3-fold defined by \eqref{K3fib} is its smoothing.
This is a generic property, one can see the toric constructions introduced in \cite{Alim:2009rf,Alim:2009bx} as a prescription how to cut a CY hypersurface into a normal crossing of Fano varieties.}

The difference between the fibration \eqref{K3fib} and the fibration of $P'$ over the $z$-plane is the additional K\"ahler modulus $\at$. For $\at=0$ the additional $\CP^1$ shrinks and the two geometries agree, $y_0y_1=0$ implies $y_0=0$ in this case. Especially the first component of the singular fiber $Z|_{y_0y_1=0}$ is $x_1=0$ and the coordinates for the second one are $(x_1:x_2:x_3:x_4)$ in both cases.

The complex structure of the fibration is singular at $y_0y_1=0$/$z=0$ and has a monodromy if we move around this point. The dual K\"ahler monodromy of $Z^*$ is a shift in the $B$-field. This signals the presence of an NS5 brane on a divisor dual to the class of the corresponding $B$-field. In the next chapter we show that this is indeed the divisor \eqref{divisor} whose relative periods obey a GKZ system with charges \eqref{K3fib}. The modulus of this divisor  is mirror to the additional K\"ahler modulus $\hat{t}$. 

We did not discuss possible $\mathcal{O}(y_0y_1)$ terms in the equation \eqref{K3P}. Such terms signal the additional freedom in the variation of the complex structure of the fiber over the base. Depending on the choice of these terms, the dual geometry is the trivial fibration $Z^*\times\IC$ or an honest fibration with a varying K\"ahler structure.

The same construction is possible for any realization of a K3 surface or for 3 or 4 dimensional CY hypersurfaces. Above we started with an elliptic K3 to make contact with the torus. But note that locally, at the vanishing locus of the $S^1$, the singular fiber always looks like the product of the degeneration locus times a cone. Mirror Symmetry in the SYZ picture always involves one T-duality in the transverse geometry, so applying Mirror Symmetry fiberwise should give rise to a dual geometry involving NS5 branes. In the following we use toric methods to show that the complex structure monodromy around the central fiber always maps to the monodromy in the $B$-field caused by an NS5 brane.

\section{Divisors and Monomials}
\label{polytopes}

First we fix the notation and repeat some facts about reflexive polytopes and associated CY hypersurfaces that we will need in the following. For more information see \cite{Batyrev:1994hm,mibo}.
$\nu_i\in \Delta^*$ are integral points of the lattice polytope $\Delta^*$ of the CY $Z^*$, $\mu_j\in \Delta$ are integral points in the dual lattice polytope $\Delta$ of $Z$. $\nu_0$ and $\mu_0$ are the unique interior points and $\braket{\nu_i}{\mu_j}=\braket{\mu_j}{\nu_i}\in \IZ$ is the natural pairing. We take the whole polytope to lie in an affine plane of distance $1$ to the origin, such that $\braket{\nu_0}{\mu_j}=1$ for all $\mu_j$ and $\braket{\mu_0}{\nu_i}=1$ for all $\nu_i$.

Taking the vectors $\nu_i-\nu_0$ as generators of one dimensional cones, we can construct the fan of the ambient space of $Z$ from $\Delta^*$ and likewise the fan of the ambient space of $Z^*$ from $\Delta$. One dimensional cones correspond to divisors $x_i=0$ of the ambient space and by restriction onto the hypersurface to toric divisors of $Z$. So there is a correspondence $\nu_i\leftrightarrow x_i=0$ and $\mu_j \leftrightarrow \x_j=0$, $i\,,j\not=0$, between integral points and divisors and we choose to label the coordinates $x$ and $\x$ with the same indices as $\nu$ and $\mu$.

Moreover all integral points $\mu_j$ of the polytope $\Delta$ correspond to a monomial $x^{\mu_j}$ in the hypersurface equation $P=0$ of $Z$ and likewise $\nu_i$ to monomials $\x^{\nu_i}$ in $\tiP=0$. Here we use the notation $x^{\mu_j}:=\prod_i x_i^{\braket{\mu_j}{\nu_i}}$ and  $\x^{\nu_i}:=\prod_j \x_j^{\braket{\nu_i}{\mu_j}}$.

The integral points $\mu_i$, $i\not=0$ correspond thus both to a monomial in the defining equation $P=0$ of $Z$ and to a toric divisor of $Z^*$. Mirror symmetry exchanges this data. Close to the large volume point in the K\"ahler moduli space of $Z^*$ and to the point of maximal unipotent monodromy in the complex structure moduli space of $Z$ this identification gives rise to the "monomial divisor mirror map" \cite{Aspinwall:1993mm}. 
A change of the K\"ahler volume of a two cycle dual to a given toric divisor is mapped to a change of the prefactor of the corresponding monomial in the hypersurface equation $P=0$ and thus to a change of complex structure. In particular, at the point of maximal unipotent monodromy this prefactor vanishes and moving around this point we get a monodromy $\tau\to \tau+1$ in the complex structure moduli space of $Z$ and $t\to t+1$ in the K\"ahler moduli space of $Z^*$.

K\"ahler classes of the ambient space\footnote{Most of these restrict to K\"ahler classes of the CY.} of $Z$ are in one to one correspondence with a certain base for the set of linear relations between points of the polytope $\Delta^*$, $\sum_i l_i^m \nu_i=0$. For this base, the entries of the charge vectors $l^m$ are the intersection numbers between a curve dual to the corresponding K\"ahler class and the divisors $x_i=0$. Divisors with the same entries for all $l^m$ and thus the same intersection numbers are equivalent and dual to the same K\"ahler class. The relation $\sum_i l_i^m \nu_i=0$ translates to the condition that all monomials $x^{\mu_j}$ of the hypersurface equation $P=0$ are in the divisor class of the anticanonical bundle. 

With the construction of chapter \ref{construction} we can choose any divisor $\mathcal{D}$ in $Z^*$ given by $\tilde{Q}=(\x^{\nu_1}+\x^{\nu_2}+..+\x^{\nu_n})/\gcf$, where $\gcf$ is the greatest common factor of the appearing monomials $x^{\nu_i}$. In the following we explain how to identify the divisor class in terms of one dimensional cones generated by $\mu_a-\mu_0$ and thus in terms of points $\mu_a$ of the dual polytope. Next we study the proposed mirror geometry and determine which monomials of $P$ depend on the base coordinate $y_1...y_n$ of the $CY$ fibration. We will see that exactly the monomials $x^{\mu_a}$ get suppressed in the central fiber over $y_1...y_n=0$, where $\mu_a$ are the points that correspond to the divisor class of $\mathcal{D}$. The monomial divisor mirror map \cite{Aspinwall:1993mm} then assures a monodromy of the complex structure in agreement with the proposed picture of a geometrization of NS5 branes by mirror symmetry.

In the simplest case we have only two monomials that determine the divisor, $\tiQ=(b_1 \x^{\nu_a}+b_2 \x^{\nu_b})/\gcf$. The divisor class can be read of either the nominator or the denominator of $\frac{\x^{\nu_a}}{\x^{\nu_b}}=\x^{\nu_a-\nu_b}$. Choosing the nominator we find the divisor $\x_1^{k_1}\x_2^{k_2}..=0$ with the multiplicities 
\begin{align}
\label{conddiv}
\begin{array}{cc}
k_j=\braket{\nu_a-\nu_b}{\mu_j} &\text{if}\;\braket{\nu_a-\nu_b}{\mu_j} >0\,,\\
k_j=0 &\text{if}\;\braket{\nu_a-\nu_b}{\mu_j} \leq 0\,.\\
\end{array}
\end{align}

The proposed mirror geometry is a CY fibration over $\IC$. Enlarging the polytope $\Delta^*$ to a polytope $\hat{\Delta}^*$ with points $(\Delta^*,0)$ and $(\nu_a,1)$, $(\nu_b,1)$ we find a new relation $\hat{l}$ between the points of $\hat{\Delta}^*$ and thus a condition on the possible monomials in the coordinates $x_i$ and $y_1$, $y_2$.\footnote{If one of the lattice points $\nu_{a/b}$ is the interior point $\nu_0$, the coordinate corresponding to $(\nu_{a/b},0)$ is $P$ and not $x_{a/b}$. This does not change the following discussion however.}
\begin{center}
\bee
\label{enlarged}\phantom{arrrgh}\eee\vskip-2.0cm
\begin{tabular}{cccccc}
 &$(\nu_a,0)$ & $(\nu_b,0)$ & ... & $(\nu_a,1)$ & $(\nu_b,1)$  \\
 &$x_a$ & $x_b$ & ... & $y_a$ & $y_b$  \\
 \hline
$\hl$ & -1 & 1 & ... & 1 & -1 
\end{tabular} ,
\end{center}

In addition we have all the relations $l^m$ between the points $\nu_i$. Imposing them gives the set of $x^{\mu_j}$ as possible monomials of the hypersurface equation $P=\sum_j a_j x^{\mu_j}=0$ of a general fiber. The additional condition forces us to multiply some of these monomials with $y_a$ or $y_b$. We get a hypersurface equation $P=\sum_j a_j(y_a,y_b) \; x^{\mu_j}$. We are interested in the behavior of the coefficients $a_j(y_a,y_b)=y_a^{k_j}y_b^{l_j}(a_j^0 +\mathcal{O}(y_ay_b))$ close to $y_ay_b=0$ so we neglect the subleading contributions $\mathcal{O}(y_ay_b)$. The monomials have to be neutral under the charges of the vector $\hat{l}$. The power of $x_a$ in $x^{\mu_j}$ is $\braket{\nu_a}{\mu_j}$ and similarly for $x_b$, so we get monomials $x^{\mu_j}y_a^{k_j}y_b^{l_j}$, where
\begin{equation}
\label{condmon}
\begin{array}{ccc}
k_j=\braket{\nu_a-\nu_b}{\mu_j},& l_j=0\,, &\text{if}\;\braket{\nu_a-\nu_b}{\mu_j} >0\,,\\
k_j=0\,,&l_j=0\,, &\text{if}\;\braket{\nu_a-\nu_b}{\mu_j} = 0\,,\\
k_j=0\,,&l_j=-\braket{\nu_a-\nu_b}{\mu_j}\,,&\text{if}\;\braket{\nu_a-\nu_b}{\mu_j} <0\,.
\end{array}
\end{equation}
By a rescaling of the $x_i$ that leaves the holomorphic $(n,0)$ form and thus the period integrals invariant it is always possible to combine $y_a$ and $y_b$ to the product $y_a y_b$. In the monomials this replaces e.g. $y_a\to y_ay_b$ and $y_b\to 1$ and we are left with $P=\sum_j (y_a y_b)^{k_j} x^{\mu_j} (a_j^0 + \mathcal{O}(y_ay_b))$. Comparing conditions \eqref{conddiv} and \eqref{condmon} we see that monomials corresponding to a point $\mu_j$ are suppressed with a power $k_j$, if the divisor $\mathcal{D}$ contains the divisor $\x_j=0$ $k_j$ times. The monomial divisor mirror map thus assures fitting monodromies.

By a different rescaling of the $x_i$ we could as well replace $y_a\to1$ and $y_b\to y_ay_b$. This would suppress monomials that started out with positive powers of $y_b$ by $(y_ay_b)^{l_j}$ . The corresponding points $\mu_j$ correspond to the divisors in the denominator of $\x^{\nu_a-\nu_b}$. This reflects the equivalence of the divisor classes.

This can be generalized to divisors $\tiQ=(b_1 \x^{\nu_a}+b_2\x^{\nu_b}...+b_n \x^{\nu_*})/\gcf$ with more than two monomials. For each additional monomial we get a new independent relation $\hat{l}_m$. Imposing one relation we multiply all monomials $x^{\mu_i}$ with new coordinates $y_*$ that correspond to divisors $\x_j=0$ that differ between two of the monomials of $\tiQ$. After imposing all relations, all monomials $x^{\mu_i}$ are multiplied by some power of some new coordinates $y_*$ up to monomials that correspond to divisors in the $\gcf$ of $\tiQ$. A rescaling of $x_i$ again collects all $y_*$ into the coordinate on the base of the fibration. In the following we give some explicit examples of divisors with several moduli.

\section{Further examples}
\subsection{Torus, charge 3}

As an example with more than two monomials in the divisor equation we again consider a torus $Z^*$, defined by $\tiP=a_1\x_1^3+a_2\x_2^3+a_3\x_3^3+a_0 \x_1\x_2\x_3=0$ in $\mathbb{P}^2/\mathbb{Z}_3$. This time we add an NS5 brane on the divisor $\tiQ=b_1 \x_1^3+b_2 \x_2^3+b_3\x_3^3+b_0 \x_1\x_2\x_3=0$, again localized at the origin of $\mathbb{C}$ and wrapping $\IR^6$. 

The dual non compact 2-fold is given by the GLSM
\begin{center}
\bee\label{Twgh}\phantom{arrrgh}\eee\vskip-1.5cm
\begin{tabular}{cc|cccccccc}
& $P$ & $x_1$ & $x_2$ & $x_3$ & $y_0$ & $y_1$ & $y_2$ & $y_3$ \\
\hline
$l$ & -3 & 1 & 1 & 1 & 0 & 0 & 0 & 0 \\
$\hl^1$ & -1 & 1 & 0 & 0 & 1 & -1 & 0 & 0 \\
$\hl^2$ & -1 & 0 & 1 & 0 & 1 & 0 & -1 & 0 \\
$\hl^3$ & -1 & 0 & 0 & 1 & 1 & 0 & 0 & -1 \\
\end{tabular} ,
\end{center}
with hypersurface
\begin{equation}
P=x_1x_2x_3+ y_0\, q^3(x_1y_1,\, x_2y_2,\, x_3y_3) + \mathcal{O}(y_0y_1y_2y_3)\,.
\end{equation}
This is a fibration of the dual torus $Z$ over $\IC$, the coordinate on $\IC$ is $y_0y_1y_2y_3$. The period integrals can be brought into the standard form by a rescaling $x_1\to x_1 \frac{(y_2y_3)^{1/3}}{y_1^{2/3}},\,x_2\to x_2 \frac{(y_1y_3)^{1/3}}{y_2^{2/3}},\,x_3\to x_3 \frac{(y_1y_2)^{1/3}}{y_3^{2/3}}$, 
\begin{equation*}
 \int\frac{\Xi}{x_1x_2x_3+y_0\, q^3(x_1y_1,\, x_2y_2,\, x_3y_3)}=\int\frac{\Xi}{x_1x_2x_3+y_0y_1y_2y_3\, q^3(x_1, x_2, x_3)}\,.
\end{equation*}
The complex structure of the fiber behaves like $\tau=3\ln(y_0y_1y_2y_3)+\mathcal{O}(y_0y_1y_2y_3)$ near $y_0y_1y_2y_3=0$. We get the factor 3 as all monomials $x_i^3$ are suppressed by $y_0y_1y_2y_3$. Alternatively, after a rescaling a single monomial e.g. $x_3^3$ is suppressed by $(y_0y_1y_2y_3)^3$. The monomials $x_i^3$ are related to the divisors $\x_i=0$ by the monomial-divisor mirror map. The class of a point is dual to the K\"ahler class, so we would indeed expect a monodromy $\rho\to\rho+3$ for an NS5 brane wrapped on the divisor $\tiQ=b_1 \x_1^3+b_2 \x_2^3+b_3\x_3^3+b_0 \x_1\x_2\x_3=0$.
\newline

\subsection{Torus, charge 1}
We already described NS5 branes on divisors of class $2[pt]$ and $3[pt]$ in a torus, but not the elementary situation of a single brane localized on one point. The toric realization of the dual geometry is a little bit more complicated but straightforward after the general discussion of section \ref{polytopes}. This time we realize the torus $T^*$ as a degree 3 hypersurface $\tiP=0$ in $\CP^3$. There are ten possible monomials $\x^{\nu_i}$ out of which we can choose two to define $\tiQ$. Usually one restricts the number of monomials by $PGL(3,\IC)$ coordinate changes and only keeps $\x_1^3$, $\x_2^3$, $\x_3^3$ and $\x_1\x_2\x_3$. In the polytope the other monomials correspond to interior points of a codimension one face. On the mirror side these points correspond to divisors in the ambient space that are not hit by the generic CY hypersurface.
However, if we want to express $\tiQ=\x_1+\x_2$ in terms of monomials $\x^{\nu_i}$ we have to use at least one of these additional points, e.g. $\tiQ=(b_1\x_1^3+b_2\x_1^2\x_2)/x_1^2$. 
The GLSM for of the dual geometry is given by
\begin{center}
\beo\phantom{arrrgh}\eeo\vskip-1.5cm
\begin{tabular}{cc|ccccccc}
& $P$ & $x_1$ & $x_2$ & $x_3$ & $x_4$ & $y_1$ & $y_2$  \\
\hline
$l^1$ & -3 & 1 & 1 & 1 & 0 & 0 & 0 \\
$l^2$ & 0 & 2 & 1 & 0 & -3 & 0 & 0 \\
 &  &  &  & $\vdots$ &  &  &  \\
$\hl$ & 0 & -1 & 0 & 0 & 1 & 1 & -1 \\
\end{tabular} ,
\end{center}
where $x_4$ is the coordinate for one of the blow-ups of the singularities of $\CP^2/\IZ^3$ and for ease of notation we omitted further blow-up coordinates and relations for them. These relations however have to be included to determine the allowed monomials for $P$.\footnote{In constructing the dual polytope one does so automatically, we focus here on the relations as we have to include the additional constraint by $\hl$.} We find 
\begin{equation*}
P=x_1^3x_4^2y_1+x_2^3x_4y_2+x_3^3+x_1x_2x_3x_4+\mathcal{O}(y_1y_2)\,.
\end{equation*}
After a rescaling of $x_i$ either the monomial with $x_1^3$ or $x_2^3$ is suppressed close to $y_1y_2=0$ and we get the expected monodromy $\tau\to\tau+1$.

\subsection{Quintic}
\label{Quintic}
For CY 3-folds, geometries of the type discussed in chapter \ref{construction} were already used in \cite{Alim:2009rf,Alim:2009bx} to calculate superpotentials. The simplest example is the mirror quintic $\tiP=\x_1^5+\x_2^5+\x_3^5+\x_4^5+\x_5^5+\x_1\x_2\x_3\x_4\x_5$ with NS5 brane on $\tiQ=\x_1^4+\x_2\x_3\x_4\x_5$. Here all additional coordinates needed to describe the blow-ups are scaled to one for ease of notation. The intersection of $\tiP$ and $\tiQ$ is a covering of a K3 surface, for more details see \cite{Alim:2009bx,Alim:2010za}. The dual quintic fibration is
\begin{center}
\beo\phantom{arrrgh}\eeo\vskip-1.5cm
\begin{tabular}{cc|cccccccc}
& $P$ & $x_1$ & $x_2$ & $x_3$ & $x_4$ & $x_5$ & $y_0$ & $y_1$ \\
\hline
$l$ & -5 & 1 & 1 & 1 & 1 & 1 & 0 & 0 \\
$\hl$ & -1 & 0 & 0 & 0 & 0 & 0 & 1 & -1 \\
\end{tabular} ,
\end{center}
with hypersurface
\begin{equation*}
P=x_1p^4(x_1y_1,\,x_2,\,x_3,\,x_4,\,x_5)+ y_0\, q^5(x_1y_1,\, x_2,\, x_3,\,x_4,\,x_5) + \mathcal{O}(y_0y_1)\,.
\end{equation*}
After a rescaling of $x_i$ all monomials without $x_1$ are suppressed by $y_0y_1$. As shown in section \ref{polytopes} this are the monomials that correspond to the divisor class of $\mathcal{D}$.

The singular locus in the central fiber is $p^4(x_2,x_3,x_4,x_5)=0$ and $x_1=\sqrt{\hat{t}}$ in $\CP^4$, where $\at$ is again the K\"ahler parameter associated with $\hl$. This is a K3 surface and it is the mirror of the K3 surface whose covering is wrapped by the NS5 brane in the mirror quintic.

\section{Conclusions}

We presented evidence that CY fibrations of the type discussed in chapter \ref{construction} can be interpreted as mirrors of CY hypersurfaces with NS5 brane on a divisor. This gives a new interpretation of recent calculations in open string mirror symmetry. \newline
A generalization to complete intersection CY manifolds should be straightforward and the idea should also carry over to other CY that were studied in open string mirror symmetry \cite{Shimizu:2010us}. The construction allows to study mirror symmetry for a pair of a CY and divisor without specifying an A-type brane on the mirror. Nevertheless the geometry should encode information of an A-type brane as discussed in \cite{Alim:2009rf,Alim:2009bx}. The role of the A-type brane is played by the degeneration locus in the singular fiber. It would be interesting to investigate such a correspondence, e.g. by a lift to M-theory.

We would like to note some observations. We saw in the Quintic example \ref{Quintic} that the degeneration locus is the mirror of the K3 surface that determines the subset of open periods. This is true also for all examples in \cite{Alim:2010za}, where these K3 "subsystems" in $Z^*$ were used to calculate numbers of disks ending on Lagrangian submanifolds in $Z$. It might be rewarding to study this "mirror symmetry" between degeneration locus and divisor in the light of the Strominger Yau Zaslow conjecture. The Lagrangian torus fibration has always one leg in a normal direction to the divisor. If the remaining directions restrict to a Lagrangian torus fibre of the divisor, one would expect the degeneration locus to be the mirror. Note that the construction is possible for any non rigid divisor.

For $d-1$ dimensional divisors with more then one modulus the degeneration locus in the dual geometry falls apart into different components that only meet in complex codimension one. As ${ dim}(H^{(d-1,0)})>1$ for such divisors this is what one would expect for the mirror geometry, there should be more then one class of points. Such a  structure appeared e.g. in \cite{Kapustin:2010aa}.

\section*{Acknowledgements}
I thank Murad Alim, Michael Hecht, Hans Jockers, Peter Mayr and Masoud Soroush for collaboration on closely related topics and Andres Collinucci, Stefan Groot Nibbelink, Michael Kay, Christian Roemmelsberger and Ahmad Zein Assi for discussions or comments. The work was supported by the Studienstiftung des deutschen Volkes.
%

\providecommand{\href}[2]{#2}\begingroup\raggedright\endgroup

\end{document}